\documentstyle[prb,aps,multicol,epsf]{revtex}

\author{S. N. Dorogovtsev\cite{E}}
\title{Nonlinear nonlocal diffusion of magnetic flux in
thin type-II superconductors and Josephson junction arrays:
Exact solutions
}
\draft
\address{A.F.Ioffe Physico-Technical Institute, 194021 St.Petersburg, Russia
}

\begin{document}

\maketitle
\begin{abstract}
An exact solution of the nonlinear nonlocal diffusion problem is obtained
that describes the evolution of the magnetic flux injected into a soft or
hard type-II superconductor film or a two-dimensional Josephson junction
array. (The magnetic field in vortices is assumed
to be perpendicular to the film; the electric field induced by
the vortex motion is proportional
to the local magnetic induction; flux creep in the hard superconductors
under consideration is
described by the logarithmic $U(j)$ dependence.) Self-similar flux
distributions with sharp square-root fronts are found. The fronts are shown
to expand with power law time-dependence. A sharp peak in the middle of the
distribution appears in the hard superconductor case.
\end{abstract}


\begin{multicols}{2}

There are two kinds of problems of magnetic flux evolution in the type-II
superconductors. In the first type problems the flux lines are
parallel to the surfaces of the superconductor plate. In this case the
motion of every vortex is determined by it's nearest environment, so
local nonlinear diffusion takes place. It can be described by nonlinear
differential equations
\cite{vinokur,schnack,blatter,brandtrev,bd1,bd2,gilchrist,kozel,wang},
which are similar to classical equations for nonlinear diffusion (see, for
example, \cite{landau,samarskii,aronson,tsallis,stariolo}).
In the second type problems the
flux lines are perpendicular to the surface of a thin superconductor\cite
{brandt1,brandt2,gurevich,dor}. Then every vortex is influenced by {\it all}
vortices in the superconductor, and one observes a so called {\it nonlocal}
nonlinear diffusion of the vortices, which can described by a complicated
integral differential equation. Note that the geometry of a flat sample in a
perpendicular magnetic field is realized in most experiments which study the
space-time image of the magnetic flux in type-II superconductors (see \cite
{duran} and references therein). Very similar to that is the problem of
the magnetic flux penetration into two-dimensional Josephson junction arrays
\cite{majh}. Macroscopic dynamical equations for the magnetic flux in the
last case are the same as for type-II syperconductor films.

Up to now the last geometry was theoretically studied only in the
particular case
of the flux flow resistivity independent of the magnetic field.
The situation where the electric field $E$ depends linearly on
the current density $j$ \cite{brandt1,brandt2} was considered
numerically. Switching on of a constant
magnetic field in the situation, in which the electric field depends
exponentially on $j$ and the flux flow resistivity is again independent of
the local magnetic induction $B$, was treated exactly in paper \cite
{gurevich}. (Note that, strictly speaking, one can introduce the quantity
''flux flow resistivity'' only if $E\propto j$. When we use this term for
brevity, we mean that $E(B,j)$ can be written in the form $E(B,j)=\rho
(B)\epsilon (j)$, where $\rho $ and $\epsilon $ are some functions.)

In the present paper we shall exactly describe the evolution of the magnetic
flux injected into the film of a type-II superconductor, in which the
flux flow resistivity is proportional to the local magnetic induction (that
is the most widespread situation \cite{larkin}). The cases of other field
dependencies of the flux flow resistivity will be discussed only briefly.
Note that the flux is assumed to be injected into a middle of the film
instead of penetration of the magnetic flux from boundaries which is
studied usually.

Let us consider the evolution of the magnetic flux injected in the infinite
thin film (the $xy$ plane) of a type-II superconductor (the flux lines are
perpendicular to the surface). We assume the problem to be
homogeneous along $y$,
so the local magnetic induction $B$ depends only on the coordinate $x$ and
on time. The current flows along $y$. An applied magnetic field is absent.
We write out immediately the equation for $B(x,t)$ in the most general form
(afterwards we shall show how it can be obtained):

\begin{eqnarray}
\label{E1}\frac{\partial B\left( x\right) }{\partial t}=D\frac \partial
{\partial x}\phantom{eeeeeeeeeeeeeeee}\nonumber\\
\times
\left\{ \left| B\left( x\right) \right| ^q\left| \int_{-\infty
}^\infty du\frac{B\left( u\right) }{u-x}\right| ^psign\left[ \int_{-\infty
}^\infty du\frac{B\left( u\right) }{u-x}\right] \right\} .
\end{eqnarray}
Here we have introduced the dimensional coefficient $D$ and the constants $%
p>0,q\geq 0$. It is convenient for us to use Eq. (\ref{E1})
rather than the equation for the current density (see \cite{brandt1,brandt2}%
) or than the equation for the electric field (see \cite{gurevich}).

Let us first show how Eq. (\ref{E1}) can be obtained for a soft
type-II superconductor.
The continuity equation for the vortex density $n$ ($%
B=\Phi _0n$, $\Phi _0$ is a flux quantum) has the form:

\begin{equation}
\label{E2}\frac{\partial n}{\partial t}+div{\bf J}_v=0.
\end{equation}
In the simplest case, the vortex flow ${\bf J}_v$ which is proportional to
the electric field, can be expressed in terms of the vortex density and the
vortex velocity: $J_v=n v$ ($v$ is the velocity of vortices; the sign of
$n(x)$ is plus if the magnetic field in vortices in the point under
consideration is parallel to the $z$ axis and it is minus otherwise;
we assume that vortices and antivortices annihilate instantaneously).
One can write for a soft superconductor: $\eta v=\left( \Phi _0/c\right) j$,
where $j$ is the current density, $c$ is the velocity of light, and the
viscosity $\eta $ is related to the normal phase resistivity $\rho _n$ by
the equation $\Phi _0/\eta =\rho _nc^2/H_{c2}$ ($H_{c2}$ is the upper
critical magnetic field). Thus we have for a soft superconductor in the
considered geometry:

\begin{equation}
\label{E3}\frac{\partial B}{\partial t}=-\frac{\Phi _0}{c\eta }\frac
\partial {\partial x}\left( \left| B\right| j\right) .
\end{equation}
This equation is also available in the case of the flux lines parallel to
the surface of a soft superconductor - see \cite{bd2}, where we discuss the
reasons for the modulus of $B$ to appear. However, in this situation
$j=-(1/4\pi c)\partial B/\partial x$, so

\begin{equation}
\label{E3a}
\frac{\partial B}{\partial t}=-\frac{\Phi _0}{4\pi c\eta }\frac
\partial {\partial x}\left( \left| B\right|
\frac{\partial B}{\partial x}\right) .
\end{equation}

In a thin infinite superconductor film of the thickness $d$, $B(x,t)$
related to the current density $j(x,t)$ by Amp\'{e}re's law, which reads

\begin{equation}
\label{E4}B\left( x\right) =\frac{2d}c\int_{-\infty }^\infty du\frac{j\left(
u\right) }{u-x},
\end{equation}
i.e. the Gilbert transformation \cite{muskhel}. It may be inverted to
give

\begin{equation}
\label{E5}j\left( x\right) =-\frac c{2\pi ^2d}\int_{-\infty }^\infty du\frac{%
B\left( u\right) }{u-x}.
\end{equation}
Inserting the expression (\ref{E5}) for $j$ into Eq. (\ref{E3}), one obtains
Eq. (\ref{E1}) with $q=p=1$ and $D=\Phi _0/(2\pi ^2d\eta )$. (The
corresponding stationary problem was considered in \cite{aslamaz}.)

Frequently, in the flux creep regime, one can assume that the effective
creep activation barrier grows logarithmically with decreasing current: $%
U(j)=U_0\ln \left| j_c/j\right| $, where $U_0$ is the characteristic scale
of the pinning energy barrier, $j_c$ is the critical current density, so the
vortex velocity $v=l_h\omega _h\left| j/j_c\right| ^{U_0/kT}j/j_c$, where $%
l_h$ and $\omega _h$ are the averaged hopping distance and
the hopping frequency
correspondingly (see \cite{vinokur} and references therein). Using again
Eqs. (\ref{E3}) and (\ref{E5}), we obtain Eq. (\ref{E1}) with $%
q=1,p=1+U_0/kT,D=l_h\omega _h(c/2\pi ^2dj_c)^p$. If before the injection of
the flux, there is the homogeneous magnetic induction $B_0$ in a sample, then
for the extra magnetic induction $\delta B(x,t)=B(x,t)-B_0\ll B_0$ one
obtains Eq. (\ref{E1}) with $q=0$, so real linearization is absolutely
impossible for hard superconductors.

Let the magnetic flux $\Phi =\int dxB(x)>0$ be injected in the sample at the
initial moment. (We assume that $B(x)$ is nonzero in a restricted range
around the point $x=0$.) At long times one may search the solution in the
following self-similar form:

\begin{equation}
\label{E5a}
B(x,t)=C_1t^{-\beta }b\left(\frac x{C_2t^\alpha}\right) ,
\end{equation}
where $C_1$ and $C_2$ are constants. Inserting this form into Eq.
(\ref{E1}) and comparing powers of $t$ in all terms of the equation, we get
the following relation for the exponents $\alpha $ and $\beta $:

\begin{equation}
\label{E6}\alpha +\beta (q+p-1)=1.
\end{equation}
Using the condition of the flux conservation $\Phi =\int dxB(x,t)=const$, we
get $\alpha =\beta =1/(q+p)$. Thus solution to Eq. (\ref{E1}) can be written
in the following self-similar form

\begin{equation}
\label{E7}B\left( x,t\right) =\left( \frac{tD}\Phi \right) ^{-1/\left(
q+p\right) }b\left( \frac x{\left( \Phi ^{q+p-1}tD\right) ^{1/\left(
q+p\right) }}\right) ,
\end{equation}
where $b(\xi )$ is defined by the equation

\begin{eqnarray}
\label{E8}-\frac 1{q+p}\frac d{d\xi }\left[ \xi b(\xi )\right] =\frac d{d\xi
}\phantom{eeeeeeeeeeeee}\nonumber\\
\times
\left\{ \left| b(\xi )\right| ^q\left| \int_{-\infty }^\infty d\zeta \frac{%
b(\zeta )}{\zeta -\xi }\right| ^psign\left( \int_{-\infty }^\infty d\zeta
\frac{b(\zeta )}{\zeta -\xi }\right) \right\} .
\end{eqnarray}
Note that a full derivative appears in the right hand part of
Eq.~(\ref{E8}) because of the equality of the exponents $\alpha $ and
$\beta $ in the regime of conservation of the injected flux.
Thus, for the values of $\xi $
where $b(\xi )$ is nonzero, we get the equation

\begin{equation}
\label{E9}\int_{-\infty }^\infty d\zeta \frac{b(\zeta )}{\zeta -\xi }%
=-\left( \frac 1{q+p}\right) ^{1/p}b^{\left( 1-q\right) /p}\left( \xi
\right) \left| \xi \right| ^{1/p}sign( \xi).
\end{equation}
(We assume that $b(-\xi )= b(\xi )$, so the integration
constant appears to be zero).

Using known analytical properties of Cauchy integral \cite{muskhel}
near the edge of a contour and the symmetry of $b(\xi )$, one may see from
Eq. (\ref{E9}) that

\begin{equation}
\label{E10}b(\xi )-b(0)=-\frac 1\pi \left( \frac 1{q+p}\right)
^{1/p}b^{\left( 1-q\right) /p}(0)\cot \frac \pi {2p}\left| \xi \right|
^{1/p}
\end{equation}
when $\xi \rightarrow 0$ (here $p>1/2$). If the function $b(\xi )$ has tails
at infinities, then one obtains immediately from Eq. (\ref{E9})

\begin{equation}
\label{E11}b(\xi )=(q+p)^{1/(1-q)}\left[ \int_{-\infty }^\infty d\zeta
b(\zeta )\right] ^{p/(1-q)}\left| \xi \right| ^{-(p+1)/(1-q)}
\end{equation}
at $\left| \xi \right| \rightarrow \infty $.

If $q<1$ (including the physically interesting case $q=0$),
there are the feature (\ref{E10}) at the center and the tails (\ref
{E11}) at $\left| \xi \right| \rightarrow \infty $. (Note that $%
\int_{-\infty }^\infty d\zeta b(\zeta )<\infty $ since $(p+1)/(1-q)>1$ for $%
q<1$.) In particular, when $q=0,p=1$, (\ref{E1}) and (\ref{E9}) turns to be
linear equations, and (\ref{E9}) has a simple solution

\begin{equation}
\label{E12}b(\xi )=\frac{const}{\xi ^2+\pi ^2}.
\end{equation}

The relation (\ref{E11}) is obviously impossible if $q\geq 1$, and in this
case $b(\xi )$ may be nonzero only near the point $\xi =0$,
so $b(\xi )=0$ if $%
\left| \xi \right| \geq \xi _0>1$. When $q\geq 1$, we shall search the
solution among such functions. It is convenient to use the function $h(\xi )$
which is nonzero in the range $-1<\xi <1$, instead of $b(\xi )$:

\begin{equation}
\label{E13}b\left( \xi \right) =\xi _0^{ 1/\left( q+p-1\right) }h\left( \xi
/\xi _0\right) .
\end{equation}
The parameter $\xi _0$ will be found at the end of our calculations. The
integral in Eq. (\ref{E9}) becomes the integral between the limits $-1$ and $%
1$. Using known analytical properties of the Cauchy integral near the edges
of the contour of integration, we see from Eq.(\ref{E9}) that $h(\xi
)\propto \sqrt{1-\xi ^2}$ near the points $\xi =\pm 1$ when $q=1$. $h(\xi
\sim \pm 1)$ can also be found in the case $q>1$. In particular,

\begin{eqnarray}
\label{E14}h(\xi )=\left( \frac 1{q+p}\right) ^{1/(p+q-1)}\left( \frac{q-1}{%
q+p-1}\right) ^{p/(p+q-1)}\nonumber\\
\times
\left( \ln \frac 1{1-\xi }\right) ^{-p/(p+q-1)}\phantom{eeeeeee}
\end{eqnarray}
when $1-\xi \ll 1$. We do not discuss this case in detail.

Let us consider the most interesting case $q=1$, for which it appears
possible to find an exact solution \cite{dor}. One may invert the Gilbert
transformation, using relations for functions restricted at the edges of a
finite interval \cite{muskhel}. Then we obtain

\begin{eqnarray}
\label{E15}
h\left( \xi \right) =\frac 1{\pi ^2}\left( \frac 1{p+1}\right) ^{1/p}
\sqrt{1-\xi ^2}\phantom{eeeeeeee}\nonumber\\
\times
\int_{-1}^1d\zeta \frac 1{\sqrt{1-\zeta ^2}}\frac{\left|
\zeta \right| ^{1/p}sign(\zeta )}{\zeta -\xi }
 =\frac 1{\pi ^2p}\left(
\frac 1{p+1}\right) ^{1/p}\nonumber\\
\times
B\left( \frac 12,\frac 1{2p}\right)
 \sqrt{1-\xi ^2}\,_2F_1\left( \frac 12-\frac 1{2p},1;\frac 32;1-\xi
^2\right) ,
\end{eqnarray}
where $B(,)$ is the beta-function and $_2F_1$ is the hypergeometric
function. $h(0)=\pi ^{-2}(p+1)^{-1/p}B\left( \frac 12,\frac 1{2p}\right) $.
Now we recall that $\Phi =\int dxB(x,t)=const$. Thus,

\begin{equation}
\label{E16}\xi _0=\pi ^{p/\left( p+1\right) }\left( p+1\right) B^{-p/\left(
p+1\right) }\left( \frac 12,\frac 1{2p}\right) .
\end{equation}
The relations (\ref{E7}), (\ref{E13}), (\ref{E15}), and (\ref{E16}) give an
exact solution of the problem under consideration in the case $q=1$.

\begin{figure}[\!h]
\epsfxsize=3.5in
\epsffile{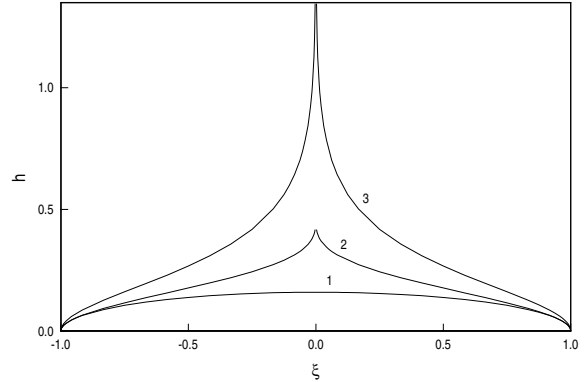 }
\caption{
\narrowtext
Scaling functions $h(\xi)$
determining the evolution of the magnetic flux injected into a type-II
superconductor film (see Eqs. (\protect \ref{E7}), (\protect \ref{E13}),
(\protect \ref{E15}), and (\protect \ref{E16})).
Curves $1,2,3$ correspond to: $q=p=1$ -- the case of a soft superconductor;
$q=1,p=3$; and $0\leq q< \infty,p\rightarrow \infty$ -- the case of a hard
superconductor.
\label{fig1}}
\end{figure}

\begin{figure}[\!h]
\epsfxsize=3.5in
\epsffile{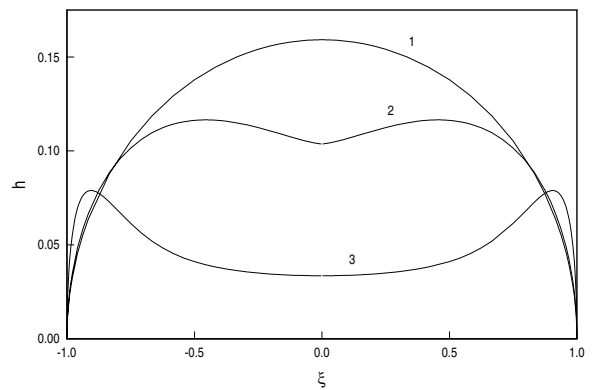 }
\caption{
\narrowtext
Same as Fig.\ \protect \ref{fig1} but curves $1,2,3$ correspond to:
$q=p=1$; $q=1,p=0.6$; and $q=1,p=0.11$.
\label{fig2}}
\end{figure}

Figs.\ \ref{fig1} and\ \ref{fig2} show scaling solutions $h(\xi )$ for $q=1$
and different $p$. The case $p\geq 1$ (Fig.\ \ref{fig1}) is interesting for
us. The solutions $h(\xi )$ with two maxima for $0<p<1$ (Fig.\ \ref{fig2})
are shown here for generality. In particular, in the case of a soft
superconductor $(q=p=1)$

\begin{equation}
\label{E17}h(\xi )=\frac{\sqrt{1-\xi ^2}}{2\pi },\quad \xi _0=2.
\end{equation}
When $q=1,p\rightarrow \infty $

\begin{equation}
\label{E18}h(\xi )=\frac 2{\pi ^2}\ln \left| \frac{1+\sqrt{1-\xi ^2}}\xi
\right| ,\quad \xi _0=\pi /2,
\end{equation}
and we obtain the solution for the case of the hard superconductor at low
temperatures. This limit has the same form also for other values $0\leq
q<\infty $. Here, in each point of the distribution, the integral $\int
d\zeta h(\zeta )/(\zeta -\xi )=const$, so we are concerned with a nonlocal
analog of the sandpile problem. As we have noted above, at $q=1$, the
solution has a square-root form near the fronts. Now we may see from Eq. (%
\ref{E15}) that

\begin{equation}
\label{E19}h(\xi \sim \pm 1)\cong \frac 1{\pi ^2p}\left( p+1\right)
^{-1/p}B\left( \frac 12,\frac 1{2p}\right) \sqrt{1-\xi ^2}.
\end{equation}

Now let us show how the current density distribution evolves.
When $q\geq 1$, one may find from the relations (\ref{E5}), (\ref{E7}), and (%
\ref{E13}) that

\begin{eqnarray}
\label{E20}j(x,t)=\frac c{2\pi ^2d}\left( \frac{tD}\Phi \right)
^{1/(q+p)}\phantom{eeee}\nonumber\\
\times
\xi _0^{1/(q+p-1)}g\left( \frac x{\xi _0\left( \Phi
^{q+p-1}tD\right) ^{1/(q+p)}}\right) ,
\end{eqnarray}
where $g(\xi )\equiv -\int_{-1}^1d\zeta h(\zeta )/(\zeta -\xi )$.; When
$\left| \xi \right| \leq 1$ and $q=1$, $g(\xi )=(p+1)^{-1/p}\left| \xi
\right| ^{1/p}sign(\xi )$. If $q=1$ and $\left| \xi \right| \geq 1$,
using (\ref{E15}) one gets

\begin{eqnarray}
\label{E21}g(\xi )=\frac 1\pi (p+1)^{-(p+1)/p}B\left( \frac 12,\frac
1{2p}\right)\nonumber\\
\times
 \frac 1\xi \,_2F_1\left( \frac 12,\frac 12+\frac 1{2p};\frac
32+\frac 1{2p};\frac 1{\xi ^2}\right)
\end{eqnarray}
\begin{figure}[\!h]
\epsfxsize=3.5in
\epsffile{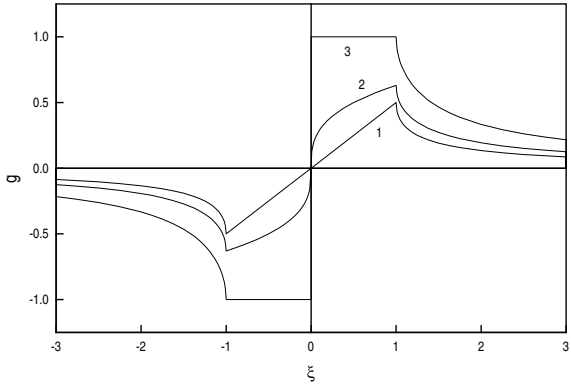 }
\caption{
\narrowtext
Scaling functions $g(\xi)$
(see Eq. (\protect \ref{E21}) for the current density distributions
corresponding to the flux distributions shown in Fig.\ \protect \ref{fig1}.
Curves $1,2,3$ correspond to: $q=p=1$; $q=1,p=3$;
and $0\leq q< \infty,p\rightarrow \infty$.
\label{fig3}}
\end{figure}
(see Fig.\ \ref{fig3}).
Near the points $\xi =\pm 1$ (at $\left| \xi \right| \geq 1$),

\begin{eqnarray}
\label{E22}
g(\xi )-g(\left| \xi \right| =1)\cong \phantom{eeeeeeeee}\nonumber\\
-\frac 1{\pi
p}(p+1)^{-1/p}B\left( \frac 12,\frac 1{2p}\right) \sqrt{\left| \xi \right| -1%
}sign(\xi ).
\end{eqnarray}
In particular, if $q=p=1$, then $g(\xi )=\xi /2$ for $\left| \xi \right|
\leq 1$ and $g(\xi )=\left[ 2\xi \left( 1+\sqrt{1-\xi ^{-2}}\right) \right]
^{-1}$ for $\left| \xi \right| \geq 1$. In the limit $p\rightarrow \infty
,q< \infty $, $g(\xi )=sign(\xi )$ for $\left| \xi \right| \leq 1$ and $%
g(\xi )=(2/\pi )\arcsin (1/\xi )$ for $\left| \xi \right| \geq 1$. In the
linear case $q=0,p=1$, as can be seen from Eq. (\ref{E12}), the current
density (\ref{E5}) is determined by the integral $-\int_{-\infty }^\infty
d\zeta b(\zeta )/(\zeta -\xi )=const\,\xi /\left( \xi ^2+\pi ^2\right) $.

Remarkably, in the studied problem, nonspecified initial distributions
of the magnetic flux evolve
to unified space-time distributions (\ref{E7}) at long times. It is
significant here that $\Phi \neq 0$. If total flux of the initial
distribution is zero, then the distribution evolves to a different
scaling form at long times.

The duration of the transition from the initial distribution of the injected
magnetic flux to the scaling solutions (see Fig.~\ref{fig1} and
Fig.~\ref{fig2})
depends on the specific form of the initial distribution.
Obviously, in the case $q \geq 1$, the initial distributions should be
localized to obtain scaling solutions at long times (see the discussion
after Eq.~(\ref{E12}).

In conclusion, let us compare the obtained solutions with the solutions of
the local equation

\begin{equation}
\label{E22a}
\frac{\partial B}{\partial t}=D\frac\partial {\partial x}
\left[ \left| B\right|
^q\left| \partial B/\partial x\right| ^p
sign\left(\frac{\partial B}{\partial x}\right)\right]
\end{equation}
describing evolution of the magnetic flux injected into a type-II
superconductor plate when the flux lines are parallel to the surfaces. (We
assume that the problem is homogeneous along a surface.)
Let the flux be injected insite of a plate. In this case,
until the vortices reach the boundaries, $B(x,t)=C_1t^{-1/(q+2p)}b\left(
x/C_2t^{1/(q+2p)}\right) $ where $C_1$ and $C_2$ are constants, and one can
see that $b(\xi )\propto \left[ 1-(\xi /\xi _0)^{(1+p)/p}\right]
^{p/(q+p-1)} $. This function has no such
striking peculiarities at the
center as in Figs.\ \ref{fig1} and \ref{fig2}.

We consider above the homogeneous along $y$ axis problem. One can also
obtain the exponents $\alpha$ and $\beta$ introduced in Eq.~(\ref{E5a})
for the situation, in which the initial distribution of the injected
flux is localized not along a line but in the vicinity of some point
on a film, so instead of the coordinate $x$ now one can introduce
the radius $r$. Then, as it can be shown, in the case of nonlinear
nonlocal diffusion in a film, $\alpha =\beta /2=1/(2q+2p-1)$ (it should be
$q+p>1/2$).
For a parallel geometry (i.e. for nonlinear local diffusion
in a plate) one can also search for scaling solutions with axial symmetry.
In this problem,
$\alpha =\beta/2=1/(2q+3p-1)$, $2q+3p>1$.

As it can be easy understood, the problem described by Eq.~(\ref{E1})
is a direct generalization of the evolution problem for
charge distribution placed on a conducting plane with dissipative
transport. Analogical problems for charges injected into a wire and
for evolution of $yz$ homogeneous charge distributions in 3D can be
also formulated. The equation for the first one is of the form

\begin{equation}
\label{E22b}
\frac{\partial n\left( x\right) }{\partial t}=D\frac \partial
{\partial x} \left\{ n(x) \int_{-\infty
}^\infty du\frac{n\left( u\right) }{|u-x|(u-x)} \right\} ,
\end{equation}
and the last one is described by the following equation,

\begin{equation}
\label{E22c}
\frac{\partial n\left( x\right) }{\partial t}=D\frac \partial
{\partial x} \left\{ n(x) \int_{-\infty
}^\infty du \, n\left( u\right) sign(u-x) \right\} ,
\end{equation}
where $n(x)$ is the charge distribution. The scaling solutions of
Eqs~(\ref{E22b}) and (\ref{E22c}) are of the form (\ref{E7}) with
$\alpha=\beta=1/2$. The scaling function $b(\xi)$ looks like

\begin{equation}
\label{E22d}
b(\xi) \propto \left( \theta(\xi+\xi_0)-\theta(\xi-\xi_0) \right)
\end{equation}
for the solution of Eq.~(\ref{E22b}), where $\theta(\xi)$ is a step
function, and

\begin{equation}
\label{E22e}
b(\xi) \propto \left( \delta(\xi+\xi_0)+\delta(\xi-\xi_0) \right)
\end{equation}
for the solution of Eq.~(\ref{E22c}). One may compare this functions
with answers for Eq.~(\ref{E1}) shown in Figs~\ref{fig1} and \ref{fig2}.
Note that an exact solution can be found generally for
the equation of the form: $\partial B/\partial t=\partial \left( \left|
B\right| F\{B\} \right) /\partial x$,
if the condition $F\{cB\}=c ^p F\{B\}$
is satisfied $(c=const\neq 0, p=const>0)$.

In summary, we have considered problem of nonlinear nonlocal diffusion of
the magnetic flux injected in an infinite thin type-II superconductor. We
have solved it exactly in the most interesting case of flow resistivity
proportional to the local magnetic induction $B$. The obtained flux
space-time distributions are of the self-similar form with rather striking
scaling functions.
Two questions remain open: how can the obtained
distributions be observed experimentally, and how do the edges of a
thin strip affect our solutions?

I wish to thank A. N. Antonov, V. V. Bryksin, A.~V.~Goltsev,
A.~M.~Monakhov, and B.~N.~Shalaev
for helpful discussions.

\end{multicols}

\end{document}